# Non-Enzymatic Graphene-Based Biosensors for Continuous Glucose Monitoring

Mohamed Serry
Department of Mechanical Engineering
The American University in Cairo
New Cairo, 11835, Egypt
mserry@aucegypt.edu

Mahmoud A. Sakr
Youssef Jameel Science and Technology Research Center
The American University in Cairo
New Cairo, 11835, Egypt

*Abstract*— A novel mediator-free, non-enzymatic electrochemical sensor, based on a graphene-Schottky junction, was fabricated for glucose detection. The sensor offers a promising alternative to the conventional enzyme-catalyzed electrochemical continuous glucose monitoring systems (CGM), as it overcomes many of the drawbacks attributed to the enzymatic nature; namely, irreversibility, drift, and interference with body fluids, which affect their accuracy, reliability and longevity. Enhanced performance of the sensors is demonstrated through the band interaction at the graphene-Schottky junction, which yields stronger forward/reverse currents in response to 50 µL glucose drop. Under optimized conditions, the linear response of the sensor to glucose concentration was valid in the range from 0 to 15 mmol/L with a detection limit of 0.5 mmol/L. The results indicated that the proposed sensor provided a highly sensitive, more facile method with good reproducibility for continuous glucose detection.

*Keywords—Glucose; Graphene; Nanosensor; Non-enzymatic; Continous Glucose Monitoring (GCM)*

## I. INTRODUCTION

Diabetes is one of the most attacking and hidden diseases all over the world. According to the IDF 387 million people living with diabetes, 46.3% undiagnosed cases, and expected to become 592 million people by 2035. Every one person over 12 is diagnosed with diabetes around the world, every 1 dollar in 9 dollars is spent on diabetes and sadly every 7 seconds one person died from diabetes and more than 21 million live births were affected by diabetes during pregnancy in 2013 [1]. Close monitoring and timely correction of elevated blood sugar can reduce the risk of diabetes-related complications. Portable glucometers, which measure blood glucose levels by finger pricks, may fail to detect rapid glucose changes. CGM sensors using electrochemical detection are intended to address this. Electrochemical detection of glucose, chemical species, and metal ions by using the oxidation reaction of molecules on the surface of the substrate have been implemented widely [2-8] to provide high sensitivity without sacrificing selectivity, however slow response time, irreversibility, and drift over time are inevitable drawbacks in these sensors. 2D material graphene showed an excellent role in the sensitivity and accuracy for various physical and chemical sensors [9].

Schottky diodes have been implemented for developing highly efficient sensors, solar cells and electrochromic devices [10]. Graphene ultrathin sheets are emerging as ideal candidates for thin-film devices and combination with other semiconductor materials such as silicon. They have been produced in the form of ultrathin sheets consisting of one or a few atomic layers directly grown by chemical vapor deposition (CVD) or by solution processing and then transferred to various substrates.

In this work, we introduce non-enzymatic, mediator free glucose sensors by growing CVD graphene on a nanostructured platinum/silicon Schottky junction (see Fig. 1). Non-enzymatic detection is achieved by the adsorption of glucose molecules on the graphene surface, glucose $OH^-$ groups interact with $O_2$ on the graphene surface. This interaction p-dopes the graphene layer, shifting the Dirac point to positive potential, thus varying the Schottky barrier's height (SBH) and width, resulting in a detectable current change.

Glucose sensing depends on the air oxidation of glucose moleulces on the surface of graphene layer and decomposition into glucono lactone and hydrogen peroxide.

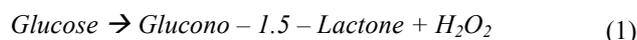

$$Glucose \rightarrow Glucono - 1.5 - Lactone + H_2O_2 \qquad (1)$$

Hydrogen peroxide molecules p ass through the porous structure of the graphene layer to the Pt catalyst where the physisorption mechanism results in $O_2$ gas, $H^+$ ions and electrons. The resulted electrons tunnel through the film towards the Si conduction band, which results in the generation of current that varies with changing the concentration of glucose in the solution. The oxidation mechanism took place in air and under controlled temperature and light exposure. Therefore, the detection method in the proposed device is non-enzymatic which provides fast response and high reversibility and usability with minor sacrifice in sensitivity and selectivity.

II. MATERIALS SYNTHESIS AND MEASURMENT PROCEDURE

*A. Platinum Nanoparticles*

It was proven that noble metals provide electrocatalytic activity to $H_2O_2$ redox reactions [11]. In our paper we focus on the advantages of Pt thin film deposited using Atomic Layer Deposition with thickness 40 nm underneath a thin film of Graphene deposited by using Chemical Vapor Deposition. Oxidation of glucose molecules on the surface of Graphene results in Hydrogen peroxide molecules which pass through the porous graphene thin film and adsorb on the platinum with a physisorption mechanism when electrochemically oxidized on the surface of platinum electrode to [12]:

$$H_2O_2 \rightarrow O_2 + 2H^+ + 2e^- \qquad (2)$$

Pt Nanoparticles are the main factor for the reduction of hydrogen peroxide molecules and the concentration of $H_2O_2$ molecules varies with the concentration of Glucose solution. The advantage of the sensor is the large surface area provided by Graphene and Pt nano particles and Pt thin film. This large surface area increases the speed of the reaction and increase the output current resulted of the oxidation of $H_2O_2$ molecules. Electrochemical activity was studied by Cyclic Voltammetry (CV) and amperometric studies. Another advantage is the stability of the device and the films after the large number of experiments. In addition the highest sensitivity which may be caused because of the structure which provides large surface area for the oxidation reactions. The sensitivity of the device was enhanced by increasing the porous structure of graphene which gives the ability to $H_2O_2$ molecules to confine and concentrate [13,14].

*B. Measurement Procedure*

D-(+)-Glucose monohydrate ($C_6H_{12}O_6 \cdot H_2O$) was obtained from Sigma Aldrich for our experiments. Glucose solution was prepared by mixing different amounts of glucose powder and dissolved in 20-µL ultrapure water. The mixture was supersonically mixed around 3 to 5 minutes. Different molarities of Glucose solutions were measured by molarity calculations [15]. The different molarities which were used in our experiments are 2, 4, 6, 8, 10, 15, and 20 mmol/L. Mixtures were prepared daily before experiments and supersonically shacked before each measurement. The sensor was connected to the 4156C high precision semiconductor parameter analyzer and we used sampling mode to measure I-t curve and sweep mode to measure I-V curve. Sensor was biased from 0.3 - 10 Volt through silicon surface and the current was measured from the Graphene terminal. All experimental work was done in dark room and fixed temperature around 21°C as the sensor is sensitive to light and heat. Drop volume was 50-µL and it was added each 60 sec of the I-t measurement. Drop size was constant in all experiments and it was added by using micropipette and it was left for 180 sec on the sensor surface and then dispensed. Sensor was cleaned with ultrapure water after every experiment and heated up around 60°C on a hotplate to evaporate ultrapure water and the traces of Glucose solution and left for seconds to cool down and then used for the nxt experiment. Urea and NaCl solutions were prepared to test device selectivity. Tests utilized Urea Extra Pure and Sodium Chloride 99.5% from Sigma Aldrich and mixed supersonically with ultrapure water for 5 min.

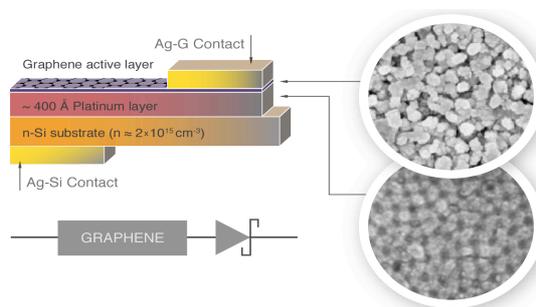

Fig. 1 Schematic of the device showing the microstructure of the graphene and platinum layers.

## III. SENSING MECHANISM

The working principle of our new device design is based on the interaction of γ–photons with the graphene layer, which changes the local electric field distribution and thus the variation of the Schottky barrier's height (SBH) and width, resulting in a detectable current change. In the case of glucose chemical sensing, when a forward-bias voltage is applied to the Schottky diode depicted earlier, i.e., an n-type Si where the work function of Pt (i.e., 5.98 eV) is greater than that of the n-Si (4.5 eV), the contact potential is reduced. Thus, the electrons move from the conduction band across the depletion region to the metal, creating a forward current (metal to semiconductor) through the junction. The current across a Pt-n-Si junction is mainly due to majority carriers. As noted above, there are three mechanisms by which current can flow: 1) diffusion of carriers, 2) thermionic emission, and 3) quantum mechanical tunneling.

By investigating the bonding attachment of glucose on a graphene surface, we will find that by rotating the carbon-carbon bonds to make the ring resemble a beach chair in shape, all of the carbon-carbon bonds are able to assume tetrahedral bonding angles. This 'chair' conformation provides the lowest possible energy conformation for cyclohexane and other six-membered rings. Moreover, as stated in [16]: "As a general rule, the most stable chair conformation of a six-membered ring will be that in which the bulkier groups are in the equatorial position."

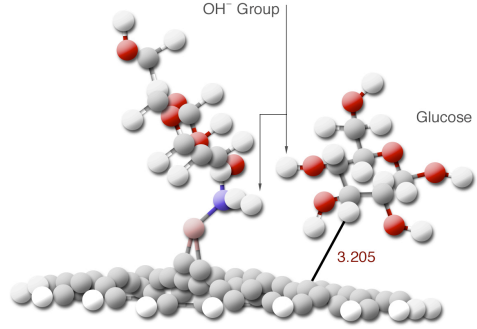

Fig. 2: Adsorption of glucose molecules on graphene surface.

Accordingly, current flowing through the junction is increased as a result of two main effects: (i) an increase in thermionic emissions due to excess electrons in the conduction band, and (ii) an increase of the tunneling current due to a decrease in the depletion layer thickness. The sensing mechanism can be summarized as follows. As glucose molecules are adsorbed at the graphene surface, $OH^-$ molecules adsorb on graphene surface. $OH^-$ p-dopes graphene layer, shifting Dirac point to positive potential. $OH^-$ molecules diffuse into the bulk of Pt metal and land on Pt/Si interface and form thin dipole layer. The barrier height is reduced by the formation of thin dipoles:

$$\Delta \Phi_b = \Delta \Phi_{b\max} C_G \quad (3)$$

where, $\Delta \Phi_b$ is the new barrier height, $\Delta \Phi_{b\max}$ is the maximum barrier height before excitation (i.e, ~ 5.98 eV in case of Pt/n-Si), and $C_G$ is the normalized parameter that is a function in glucose concentration, which can be determined experimentally. As a result, a net current flow can be determined according to:

$$I_{net} = I_{fwd} - I_{rev} \quad (4)$$

where, $I_{fwd}$ is the forward current from the substrate to graphene direction mainly due to the thermionic effect, and $I_{rev}$ is the reverse current flow in the direction from graphene to substrate due to electrons tunneled through the ultrathin platinum layer to the conduction band of the substrate.

## IV. RESULTS AND DISCUSSION

X-ray Photo-electron Spectroscopy (XPS) characterization data were performed on a K-Alpha XPS from Thermo Scientific in the range of 1 to 1350 eV to inspect the surface chemistry of the Graphene. Raman spectra of the graphene thin film were obtained by Enwave Optoronics Raman Spectroscope (λex = 532 nm, P = 500 mW, acquisition time = 10s). Atomic Force Microscopy (AFM) from Bruker, and Field Emission Scanning Electron Microscopy (FE-SEM) from Zeiss were used for structural and morphological characterizations. The SEM images of the top view of graphene layer after six months of usage is presented in Fig. 3.

Electrochemical activity was studied by Cyclic Voltammetry (CV) and the amperometric. The oxidation of glucose molecules on the surface of graphene thin layer and the porous structure of the film gives the ability for the $H_2O_2$ molecules to adsorb on the surface of the Pt layer.

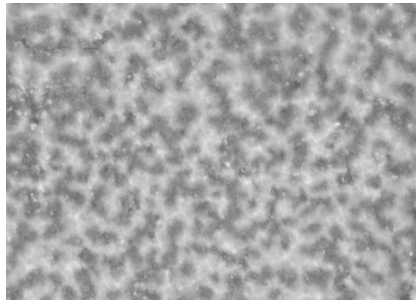

Fig. 3 SEM image of the graphene layer after 6 months of implementation as glucose sensor.

The porous structure gives the advantage of the large surface area and the bias of the diode increases the sensitivity of the sensor toward the small difference between concentrations. The sensor was connected to the 4156C high precision semiconductor parameter analyzer and we used sampling mode to measure I-t curve and sweep mode to measure I-V curve.

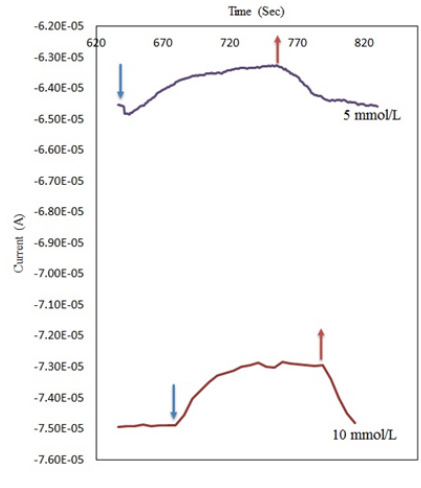

Fig. 4 Test of sensitivity, showing the current vs time curve for 5 and 10 mmol/L at 0.3V bias. Sensitivity is 0.2 µA.mmol$^{-1}$L$^{-1}$ at 0.3 V bias. Red arrows indicate when the glucose drop is added; blue arrow indicated when the glucose drop is removed response time ~60 sec.

The sensor was biased in the sampling mode with 0.3 V on the Si terminal. Different concentrations were prepared before the experimental work. Fig. 4 illustrates the I-t curve of the sensor and the different concentrations measured and the sensitivity of the sensor. The repeatability of the sensor was measured by using the sensor for more than 10 measurements of 10 mmol/L continuously without the need to clean or discharge the sensor as illustrated in Fig. 5.

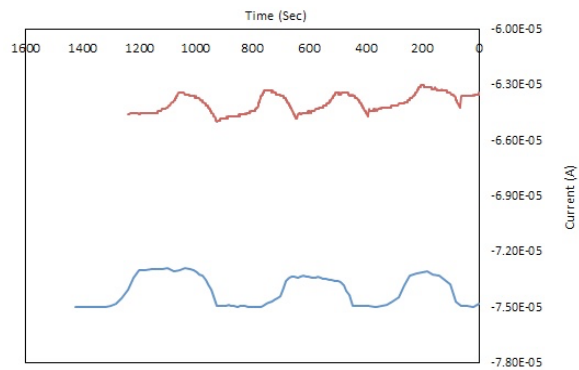

Fig. 5 Test of repeatability, showing 3 continous cycles in response to 10 mmol/L glucose for the blue line and 5 mmol/L in the red line at 0.3 V bias. Standard deviation is ~0.17 µA.mmol$^{-1}$L$^{-1}$.

The repeatability of the sensor proves that the sensor could be used for several times not a single usage and don't need to be cleaned after every measurement. Device sensitivity is 0.2 µA.mmol$^{-1}$L$^{-1}$ at 0.3 V. Standard deviation of the sensor is 0.17 µA.mmol$^{-1}$ L$^{-1}$. The advantage of incorporating the graphene layer is demonstrated clearly in Fig. 6, in which it's clear that sensitivity is enhanced tremendously in the graphene incorporated devices as compared to bare platinum surface.

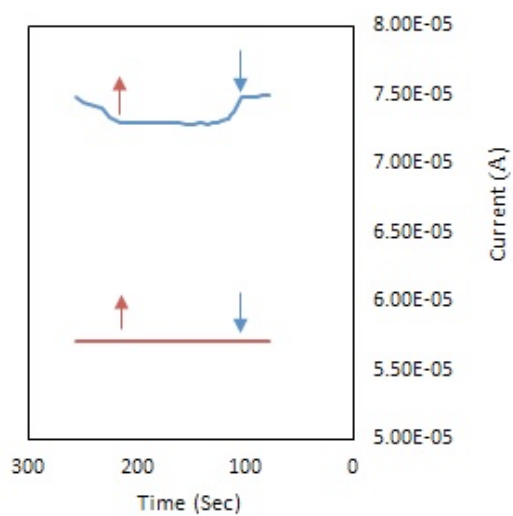

Fig. 6 Test of sensitivity of the sensor materials to 50 uL of 10 mmol/L with bias 0.3 V. Blue curve represents the response of Graphene layer and the red curve represents the response of Pt layer.

Finally, the response over wide range of bias demonstrated high linearity in the range from 7 to 9 V bias. This makes the device suitable for operation on 1.5 V coin batteries connected in series.

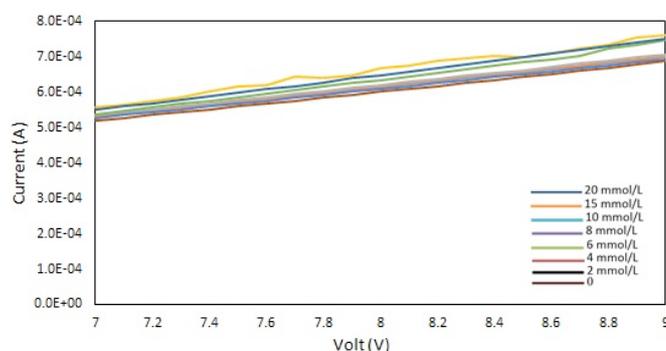

Fig. 7 Device I-V response at different glucose concentrations. Linearity over the range from 0-15 mmol/L.

ACKNOWLEDGMENT

This research was supported by an American University in Cairo (AUC) faculty support research grant.

REFERENCES

[1] Aguiree, Florencia, et al. "IDF diabetes atlas." (2013).
[2] A. Sharaf, A. Gamal, M. Serry, "High performance NEMS ultrahigh sensitive radiation sensor based on platinum nanorods capacitor," IEEE Sensors 2013, Baltimore, USA, November 3-6, 2013.
[3] M. Serry, A. Gamal, M. Shaban, A. Sharaf, High sensitivity optochemical andelectrochemical metal ion sensor, Micro & Nano Letters 8 (11) (2013)775–778.
[4] M. Serry, A. Rubin, M. Ibrahem, S. Sedky, Silicon Germanium as a novel maskfor silicon deep reactive ion etching, J. Microelectromechanical Syst. 22 (5)(2013) 1081–1088.
[5] M. Shaban, A.G.A. Hady, M. Serry, A New Sensor for Heavy Metals Detection inAqueous Media, IEEE Sensors J. vol.14 (2) (2014) 436–441, http://dx.doi.org/10.1109/JSEN.2013.2279916.
[6] A.H. Sharaf, A. Gamal, M. Serry, High performance NEMS ultrahigh sensitiveradiation sensor based on platinum nanorods capacitor, IEEE SENSORS (3-6Nov) (2013) 1–4, http://dx.doi.org/10.1109/ICSENS.2013.6688421.
[7] M. Lopez-Garcia, Y.-L.D. Ho, M.P.C. Taverne, L.-F. Chen, M.M. Murshidy, A.P.Edwards, M.Y. Serry, A.M. Adawi, J.G. Rarity, R. Oulton, Efficient out-couplingand beaming of Tamm optical states via surface plasmon polariton excitation,Applied Physics Letters 104 (no. 23) (2014) 231116.
[8] Ahmed, M. Serry, Y. H. Lee, C. B. Park, and N. Atalla. "Effect of nanoclay on the microcellular structure and morphology of high internal phase emulsion (HIPE) foams." Asia-Pacific Journal of Chemical Engineering 4, no. 2 (2009): 120-124.
[9] A. Sharaf, A. Gamal, M. Serry, "New nanostructured Schottky diode gamma-ray radiation sensor," Procedia Engineering, Vol. 87, 2014, pp. 1184-1189.

[10] M. Serry, A. Sharaf, A. Emira, A. Abdul-Wahed, A. Gamal, "Nanostructured graphene−Schottky junction low−bias radiation sensors," Sensors & Actuators: A. Physical 232 (2015) 329–340, http://dx.doi.org/10.1016/j.sna.2015.04.031

[11] Jou, Amily Fang-ju, Nyan-Hwa Tai, and Ja-an Annie Ho. "Gold Nanobone/Carbon Nanotube Hybrids for the Efficient Nonenzymatic Detection of H2O2 and Glucose." Electroanalysis 26.8 (2014): 1816-1823.

[12] Zhang, Meng, et al. "Highly sensitive glucose sensors based on enzyme-modified whole-graphene solution-gated transistors." Scientific reports 5 (2015).

[13] Jian Liu, XiangjieBo , ZhengZhao , LipingGuo. "Highly exposed Pt nanoparticles supported on porous graphene for electrochemical detection of hydrogen peroxide in living cells." Biosensors and Bioelectronics (2015).

[14] Wu, Hong, et al. "Glucose biosensor based on immobilization of glucose oxidase in platinum nanoparticles/graphene/chitosan nanocomposite film." Talanta 80.1 (2009): 403-406.

[15] Molarity Calculator and Normality Calculator for Acids and Bases." Sigma-Aldrich. Web. 6 July 2015.

[16] UC Davis Organic Chem: http://chemwiki.ucdavis.edu/Organic_Chemistry/Organic_Chemistry_With_a_Biological_Emphasis/Chapter_03%3A_Conformations_and_Stereochemistry/Section_3.2%3A_Conformations_of_cyclic_organic_molecules> [Accessed 31 January 2015].